\documentclass[twocolumn,showpacs,preprintnumbers,amsmath,amssymb]{revtex4}

\usepackage{amsmath}
\usepackage{amsfonts}
\usepackage{amssymb}
\usepackage{graphicx}
\usepackage{subfigure}
\usepackage{graphicx}
\usepackage{dcolumn}
\usepackage{bm}

\begin{document}


\title{$\gamma$ Parameter and Solar System constraint in Chameleon Brans Dick Theory }

\author{Kh. Saaidi}
 \email{ksaaidi@uok.ac.ir}
 \author{A. Mohammadi}
  \email{abolhassan.mohammadi@uok.ac.ir}
\author{H. Sheikhahmadi}%
 \email{h.sh.ahmadi@uok.ac.ir}
\affiliation{ %
Department of Physics, Faculty of Science, University of Kurdistan,  Sanandaj, Iran\\
}%


\date{\today}

\def\br{\biggr}
\def\bl{\biggl}
\def\Br{\Biggr}
\def\Bl{\Biggl}
\def\be\begin{equation}
 \def\ee{\end{equation}}
\def\bea{\begin{eqnarray}}
\def\eea{\end{eqnarray}}
\def\f{\frac}
\def\n{\nonumber}
\def\l{\label}



\begin{abstract}
The post Newtonian parameter is considered in the Chameleon-Brans-Dicke frame. In the first step, the general form of this parameter and also effective gravitational constant is obtained. An arbitrary function for $f(\Phi)$, which indicates the coupling between matter and scalar field, is introduced to investigate validity of solar system constraint. It is shown that Chameleon-Brans-Dicke model can satisfy the solar system constraint and gives us an $\omega$ parameter of order $10^4$, which is in comparable to the constraint which has been indicated in {\cite{17-a}}.
\end{abstract}

\maketitle

\newpage


\section{Introductions}
Recent cosmological and Astrophysical observations indicate that
our universe is in accelerate expansion phase. These observations
include type Ia supernovae data{\cite{1}}, Wilkinson Microwave
Anisotropic Probe (WMAP) {\cite{2}}, X-ray {\cite{3}}, large scale
structure {\cite{4}} and ete. The combine analysis of these
observations also suggest that our universe is spatially flat, and
consist $\% 73$ dark energy, $\% 23$ dark matter and remanent
matter is baryons {\cite{5}}. Although the nature and origin of
dark energy are unknown for researcher up to now; however people
have introduced some proposals to describe it. Amongst the
various proposals of dark energy to describe accelerate expansion
of universe is a tiny positive time-independent cosmological
constant, $\Lambda$, for which has the equation of state
$\omega=-1$ {\cite{6, 7}}. It is well-known the cosmological
constant scenario has two problems; cosmological constant problem
(why the cosmological constant is very smaller than its natural
expectation?), and coincidence problem (why the dark energy and
dark matter are comparable now?) {\cite{8}}. Another alternative
candidate for dark energy is the dynamical dark energy proposal.
This scenario can be originate from scalar tensors concept.
Whereas the Einstein's general theory of gravity is a geometrical
theory of space-time. The fundamental building block is a metric
tensor field. So, the theory may be called a tensor theory. This
scenario can provide a frame work within which to model
space-time variations of the Newtonian gravitational constant,
$G$. The scalar tensor theory was conceived originally by Jordan,
who started to embed a four-dimensional curve manifold in five
dimensional flat space- time {\cite{9}. The scalar tensor theories
feature a  scalar field, $\phi$, which has non-minimal coupling
with together the geometry in the gravitational action where
Brans and Dicke (BD) have introduced it {\cite{10}}. Brans-Dicke
theory or its modifications have already proved to be useful in
producing clues to the solutions for some of the outstanding
problems in
cosmology {\cite{11, 12}}.\\
The same mechanism that creates a scalar field non-minimal
coupling to the geometry in these proposals can also lead to a
coupling between the scalar and matter field. For instance two
kind of scalar fields have this condition (coupling between
scalar field and matter field) quintessence and chameleon
{\cite{13, 14}}. The quintessence mechanism is a light scalar
field mass which couples to matter directly with gravitational
strength, lead to undesirable large violation of the equivalence
principle. Therefore authors {\cite{15}} have introduced a scalar
field which it has a coupling to matter of order one, named
chameleon mechanism. Chameleons  are scalar fields whose mass
depends on the environment mass density. Indeed the chameleon
proposal produces a way to an effective mass to a light scalar
field via field self interaction, and interaction between matter
field and scalar field. Therefore the Brans-Dicke action with
non-minimal coupling between matter scalar field so called
Chameleon-Brans-Dicke model {\cite{16}}.
 Recently, Chameleon-Brans-Dicke has been used for description some models such as holographic dark energy (HDE) {\cite{16-a}}, agegraphic dark energy(ADE), and new agegraphic dark energy (NADE). All of these investigations have been worked out in the large scale, and the authors could obtain good results. Studying these works and also the works of Moffat et al. {\cite{17}}, where they have studied PPN-parameter in Jordan-Brans-Dicke cosmology, and recent work of Perivolaropoulos {\cite{17-a}} motivated us to consider constraints of solar system in Chameleon-Branes-Dicke model.
The motivation of this work is investigation the weak field limit and  solar system constraints on parameterized post Newtonian (or Edington-Robertson) $\gamma$-parameter (PPN) in Chameleon-Brans-Dicke framework. In this case the scalar has non-minimal coupling with both of matter and geometry sector. However we are going to see how the Chameleon-Brans-Dicke theory fits the observation. Since the Chameleon-Brans-Dicke theory is a metric theory, motion of matter object can be explained in the frame work of PPN approximation.\\
This work is organized in four sections, of which this introduction is the first. In section two, the action is introduced, and the field equation and the scalar field equation of motion are obtained, then we linearized these equations and find out the general solution for $\gamma$-post Newtonian parameter. In section three, we investigate the results for two special examples. In the first example, we suppose an exponential function for $f(\Phi)$, and the results are considered either in the presence of potential and in the absence of that. In the second example, the interaction between matter and scalar field is ignored, and we clearly see that our results are in perfect match with the results of previous works. Finally section four, which is conclusion, we summarize the results of this work.


\section{General Framework}
For our investigation, we consider the Brans-Dicke-
action as

\begin{equation}\label {1}
S=\int d^4x\sqrt{-g} \left( \Phi R -
\frac{\omega}{\Phi}\partial_{\mu}\Phi \partial_{\nu} \Phi -
V(\Phi) + f(\Phi)\mathcal{L}_m \right),
\end{equation}
where $g$ is the metric determinant, $R$ is the Ricci scalar
constructed from the metric $g_{\mu \nu}$, and $\Phi R$ has been
replaced with the Einstein-Hilbert term is such a way that
$G_{eff}^{-1}=16\pi\Phi$ {\cite{17-b}}, $\Phi$ is the
Chameleon-Brans-Dicke- scalar field, $\omega$ is the dimenssionless
Chameleon-Brans-Dicke constant. $V(\Phi)$ is the potential; note
that since the scalar field is dependence on the local mass we
need a monotically decreasing function for potential where
exhibit self interaction. Therefore we consider power law
potential as

\begin{equation}\label {2}
V(\Phi)=M^4 \left( \frac{M^2}{\Phi}\right)^n,
\end{equation}
we refer the reader for more review to {\cite{18}}. The last term
on the right hand said of (\ref{1}), namely
$f(\Phi)\mathcal{L}_m$, indicates non-minimal coupling between
scalar field and matter, where $f(\Phi)$ is an arbitrary function
of $\Phi$. One can obtain the gravitational field equation by
taking variation of the action (\ref{1}) with respect to the
metric $g_{\mu \nu}$ as
\begin{widetext}
\begin{equation}\label {3}
\Phi(R_{\mu \nu}-\frac{1}{2}g_{\mu \nu}R)= f(\Phi)T_{\mu
\nu}+\frac{\omega}{\Phi}(\partial_{\mu} \Phi \partial_{\nu} \Phi
- \frac{1}{2}g_{\mu \nu}(\partial_{\alpha}\Phi)^2) 
 + [\nabla_{\mu} \nabla_{\nu} -g_{\mu \nu}\Box]\Phi -
g_{\mu\nu}\frac{V(\Phi)}{2}
\end{equation}
\end{widetext}
where $T_{\mu \nu}$ indicates the energy-momentum tensor. Taking
variation of action with respect to the scalar field $\Phi$ gives
us scalar field equation of motion as

\begin{equation}\label {4}
(2\omega+3)\Box \Phi= T(f(\Phi)-\frac{1}{2}\Phi f_{,\Phi}) +
(\Phi V_{,\Phi}(\Phi)-2V(\Phi)).
\end{equation}
In this work, we consider weak field approximation. Since we are
in solar system, a weak field solution plays the same important
role that the corresponding solutions fills in general
relativity. So that we expand around a
constant-uniform background filed $\Phi_0$, and a Minkowskian
metric tensor $\eta_{\mu \nu}=$diag$(-1,1,1,1)$ as

\begin{equation}\label {5}
\Phi=\Phi_0+\phi,
\end{equation}

\begin{equation}\label {6}
g_{\mu\nu}=\eta_{\mu\nu}+h_{\mu\nu},
\end{equation}
where $h_{\mu\nu}$ is computed to the linear first approximation
only. The linearized solution for Eq.(\ref{3}) is as

\begin{widetext}
\begin{equation}\label {7}
-\frac{\Phi_0}{2}\Big[\Box(h_{\mu\nu}-\frac{1}{2}\eta_{\mu\nu}h)\Big]
= f(\Phi)T_{\mu\nu}+\partial_{\mu}\partial_{\nu}\phi -
\eta_{\mu\nu} \Box \phi - (\eta_{\mu\nu}+h_{\mu\nu})V(\Phi)
\end{equation}
and we have below relation for scalar field equation of motion in
the weak field

\begin{equation}\label {8}
(2\omega+3)\Box \phi=T(f(\Phi)-\frac{1}{2}\Phi f(\Phi)) + (\Phi
V_{,\Phi}(\Phi)-2V(\Phi))=\xi T + \zeta.
\end{equation}
\end{widetext}
Now, we expand the potential function (\ref{2}) around $\Phi_0$,
so

\begin{equation}\label {9}
V(\Phi)=\frac{M^{4+2n}}{\Phi_0^n}\left( 1-\frac{n\phi}{\Phi_0}
\right)
\end{equation}
Also, by the same algebraic analysis we have

\begin{equation}\label {10}
\xi = (f_0-\frac{1}{2}\Phi_0 f'_0)+\frac{\phi}{2}(f'_0-\Phi_0
f''_0)=C_1+C_2\phi,
\end{equation}

\begin{equation}\label {11}
\zeta=-(n+2)M^4 \left( \frac{M^2}{\Phi_0} \right)^n \left( 1-
\frac{n\phi}{\Phi_0} \right)=-C_3\left( 1- \frac{n\phi}{\Phi_0}
\right).
\end{equation}
where $f_0=f(\Phi_0)$, $f'_0$ and $f''_0$ are the first and
second derivative of $f(\Phi)$ with respect to scalar field
$\Phi$, respectively. Since our approximation related to the solar
system, we consider a stationary solution correspondence to a
gravitational mass such as earth. Substituting above results in the
field equation (\ref{7}) and scalar field equation of motion
(\ref{8}) we arrive at the differential equation for $\phi$ and
$h_{\mu\nu}$ as

\begin{equation}\label {12}
\nabla^2\phi-\frac{C_3}{\Phi_0(2\omega+3)}\phi= -
\frac{C_1+C_2\phi}{(2\omega+3)}\rho - C_3,
\end{equation}

\begin{equation}\label {13}
\nabla^2H_{00}-\frac{2M^6}{\Phi_0^2}H_{00}= -(f_0+\phi
f'_0)\rho+\frac{2M^4}{\Phi_0},
\end{equation}

\begin{equation}\label {14}
\nabla^2H_{ij}-\frac{2M^6}{\Phi_0^2}H_{ij}= -\delta_{ij} \left(  (f_0+\phi f'_0)\rho
+\frac{2M^4}{\Phi_0}\right) ,
\end{equation}
where $H_{00}=\Phi_0 h_{00}-\phi$ and $H_{ij}=\phi+\Phi_0
h_{ij}$, and note that we have taken $n=1$ . These equation are
consistent with the results of {\cite{17-a,20}} even though
there is a little difference in constants. This differences are
resulted from non-minimal coupling between matter and scalar
field and the sort of potential which we have selected. Assuming
$\rho=M_e \delta(r)$ we obtain the relations as

\begin{equation}\label {15}
\phi(r)=\frac{  \frac{C_1M_e}{4\pi (2\omega+3)r} e^{-kr} }{1-
\frac{C_2M_e}{4\pi (2\omega+3)r} e^{-kr}},
\end{equation}

\begin{equation}\label {16}
h_{00}=\frac{M_e e^{-k'r}}{4\pi\Phi_0r} \Bigg\{ f_0 + \left(
\frac{M_e f'_0e^{-k'r}}{4\pi r} + 1
\right)\frac{\frac{C_1}{2\omega+3}e^{-(k-k')r}}{1-\frac{C_2M_e e^{-kr}}{4\pi(2\omega+3)r}}
\Bigg\},
\end{equation}

\begin{equation}\label {17}
h_{ij}=\delta_{ij} \frac{M_e e^{-k'r}}{4\pi\Phi_0r} \Bigg\{ f_0 + \left(
\frac{M_e f'_0e^{-k'r}}{4\pi r} - 1
\right)\frac{\frac{C_1}{2\omega+3}e^{-(k-k')r}}{1-\frac{C_2M_e
e^{-kr}}{4\pi(2\omega+3)r}} \Bigg\},
\end{equation}
where $k=\sqrt{\frac{3}{2\omega+3}}\frac{M^3}{\Phi_0}$ and
$k'=\frac{\sqrt{2}M^3}{\Phi_0}$. Note that there is some constant
in the above relation related to the integration constant and the
last term on the right hand side of the differential equation, but
they can be omitted from the above relation. Also $h_{\mu\nu}$ is
interpreted as gravitational potential, so the presence of
constant is not so important because the potential difference is
a physical quantity in physics. From the standard expansion of
metric, namely

\begin{equation}\label {18}
g_{00}=-1+2u,
\end{equation}

\begin{equation}\label {19}
g_{ij}=(1+2\gamma u)\delta_{ij}.
\end{equation}
where $u$ is the Newtonian potential. The $\gamma$-post Newtonian
parameter can be expressed as the ratio of $(ii)$ and $(00)$
component of $h_{\mu\nu}$

\begin{equation}\label {20}
\gamma=\frac{f_0 + \left( \frac{M_e f'_0e^{-k'r}}{4\pi r} - 1
\right)\frac{\frac{C_1}{2\omega+3}e^{-(k-k')r}}{1-\frac{C_2M_e
e^{-kr}}{4\pi(2\omega+3)r}}}{f_0 + \left( \frac{M_e
f'_0e^{-k'r}}{4\pi r} + 1
\right)\frac{\frac{C_1}{2\omega+3}e^{-(k-k')r}}{1-\frac{C_2M_e
e^{-kr}}{4\pi(2\omega+3)r}}}
\end{equation}
As, it has been said in the first of this section, the effective
gravitational constant can be expressed by scalar field as

\begin{eqnarray}\label {21}
G_{eff}&=&\frac{1}{16\pi\Phi}=\frac{1}{16\pi\Phi_0}\left(
1-\frac{\phi}{\Phi_0} \right)\nonumber \\
 &=&\frac{1}{16\pi\Phi_0}\left( 1- \frac{  \frac{C_1M_e }{4\pi\Phi_0 (2\omega+3)r} e^{-kr} }{1-
\frac{C_2M_e}{4\pi (2\omega+3)r} e^{-kr}} \right) .
\end{eqnarray}
Up to now, we could attain the $\gamma$-post Newtonian parameter
and effective gravitational constant in the general form. In the
next section, we consider some special cases and compare our
result with the results of previous work.


\section{Typical Example}
In the previous section we could obtain the main relation in the
general form, now we try to consider these relation in the
special cases and compare the results with the previous work.


\subsection{First Example}
In this subsection we choose a specific function for $f(\Phi)$ as

\begin{equation}\label {22}
f(\Phi)= \exp(\frac{\Phi}{\Phi_0}).
\end{equation}
By making use of the definition, the constants $C_1$ and $C_2$ are obtained as$$C_1=\frac{1}{2}e^1
\qquad \mathrm{and} \qquad C_2=0.$$ Now we investigate the results of this
case for $M=0$ and $M\neq 0$.

\begin{itemize}

\item{\bf If $M=0$}\\
When we take $M=0$, it means that the potential have been
omitted. For this case, as well as previous constants, the
constant $C_3$ vanishes. So with attention to the definitions of
$k$ and $k'$, for scalar field $\phi$ and components of
$h_{\mu\nu}$, we have

\begin{equation}\label {23}
\phi(r)=\frac{e^1M_e}{8\pi(2\omega+3)r}
\end{equation}

\begin{equation}\label {24}
h_{00}=\frac{M_e e^1}{4\pi\Phi_0 r}\Bigg\{
1+\frac{1}{2(2\omega+3)}\left( \frac{M_e e^1}{4\pi\Phi_0 r} + 1
\right) \Bigg\},
\end{equation}

\begin{equation}\label {25}
h_{ij}=\delta_{ij} \frac{M_e e^1}{4\pi\Phi_0 r}\Bigg\{
1+\frac{1}{2(2\omega+3)}\left( \frac{M_e e^1}{4\pi\Phi_0 r} - 1
\right) \Bigg\}.
\end{equation}
And for $\gamma$-post Newtonian parameter and effective
gravitational constant we have respectively

\begin{equation}\label {26}
\gamma=\frac{(4\omega+5)+\frac{M_e e^1}{4\pi\Phi_0
r}}{(4\omega+7)+\frac{M_e e^1}{4\pi\Phi_0 r}}
\end{equation}

\begin{equation}\label {27}
G_{eff}=\frac{1}{16\pi\Phi_0}\left( 1-\frac{M_e
e^1}{8\pi\Phi_0(2\omega+3)r} \right).
\end{equation}


\item{\bf If $M\neq 0$}\\
Here, in contrast to the previous case we assume that we have
potential, so one can obtain the scalar field function and the
component of $h_{\mu\nu}$ as follow
\begin{equation}\label {28}
\phi(r)=\frac{M_e}{8\pi(2\omega+3)r} e^{1-kr}
\end{equation}

\begin{equation}\label {29}
h_{00}=\frac{M_e}{4\pi\Phi_0 r} e^{1-k'r} \Bigg\{ 1+ \left(
\frac{M_e}{4\pi\Phi_0 r}e^{1-k'r}+1 \right)
\frac{e^{-(k-k')r}}{2(2\omega+3)} \Bigg\},
\end{equation}

\begin{equation}\label {30}
h_{ij}=\delta_{ij} \frac{M_e}{4\pi\Phi_0 r} e^{1-k'r} \Bigg\{ 1+ \left(
\frac{M_e}{4\pi\Phi_0 r}e^{1-k'r}-1 \right)
\frac{e^{-(k-k')r}}{2(2\omega+3)} \Bigg\}.
\end{equation}
Now one can easily obtain the $\gamma$-post Newtonian parameter
and effective gravitational constant respectively as

\begin{equation}\label {31}
\gamma=\frac{1+ \left( \frac{M_e}{4\pi\Phi_0 r}e^{1-k'r}-1 \right)
\frac{e^{-(k-k')r}}{2(2\omega+3)}}{1+ \left(
\frac{M_e}{4\pi\Phi_0 r}e^{1-k'r}+1 \right)
\frac{e^{-(k-k')r}}{2(2\omega+3)}}
\end{equation}
and

\begin{equation}\label {32}
G_{eff}=\frac{1}{16\pi\Phi_0}\left( 1-\frac{M_e
e^{1-kr}}{8\pi\Phi_0(2\omega+3)r} \right).
\end{equation}
In this step, one can estimate the value of background scalar field, namely $\Phi_0$ from effective gravitational constant relation. According to the relation (\ref{32}), the second term in the parenthesis is the modified term for the gravitational constant. So the gravitational constant is equal to $$G_0=\frac{1}{8\pi M_{pl}^2}=\frac{1}{16\pi \Phi_0}.$$ As it is clear, $M_{pl}=2.44\times 10^{18}GeV$ {\cite{20-a}}. Therefore, the value of $\Phi_0$ can easily be estimated as order of $10^{36}$.
\begin{figure*}
\includegraphics{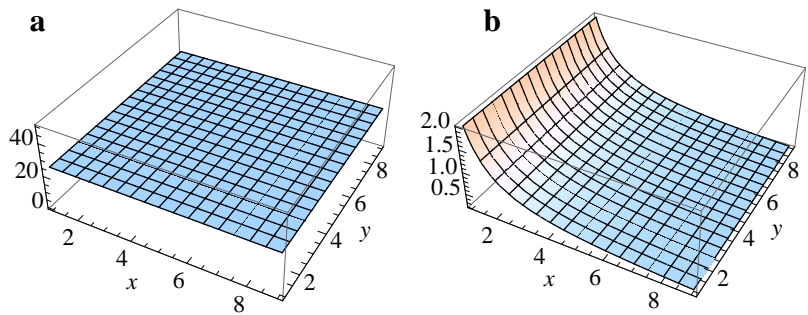}
\caption{\label{fig:epsart} The $\omega$ parameter and $G_{eff}$, effective gravitational constant, have been plotted versus $\Phi_0$ and $M_{AU}$, respectively (a) and (b) (where $\Phi_0=10^{36}x$, $m_{AU}=10^{-27}y$, $\omega=10^4 z$, and $G_{eff}=10^{-38}G$). In this case, we set $\gamma=1-0.2\times 10^{-5}$. We clearly see that $\omega$ takes the values of order $10^5$, and $G_{eff}$ tends to the values of order $10^-39$, the same order of Newtonian gravitational constant $G_0$. }
\end{figure*}
\begin{figure*}
\includegraphics{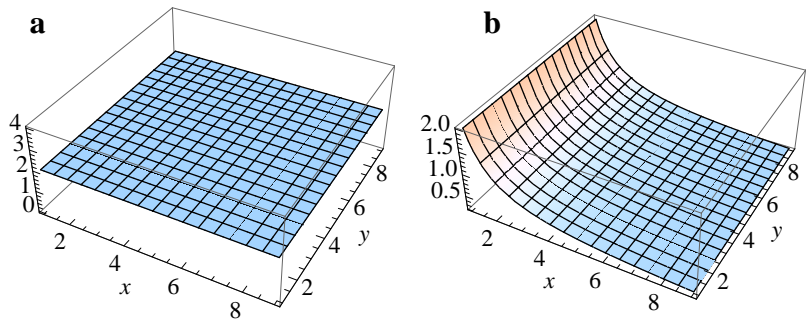}
\caption{ The plot on the left hand, (a), is realted to the $\omega$ parameter and the plot on the right hand, (b), is related to the effective gravitational constant, $G_{eff}$. For this case, we set $\gamma=1-2.5\times 10^{-5}$, and as we see, the $\omega$ parameter takes the values of order $10^4$. The effective gravitational constant is as same to the previous case, because the modified term, on the parenthesis of relation (\ref{32}), is so small.}
\end{figure*}
Now, we try to consider allowed values of $\omega$ for
$\gamma_{obs}-1=(2.1\pm 2.3)\times 10^{-5}$ \cite{17-a,18-a} with the help of
(\ref{31}). After some algeraic analysis, $\omega$ can be written
as
\begin{widetext}
\begin{equation}\label {32-a}
\omega=\frac{1}{4} \left( \frac{2\Phi_0 m_{AU}}{-\sqrt{6}M^3}
W\left[  \frac{-\sqrt{6}M^3}{2\Phi_0 m_{AU}} \Bigg\{
\frac{A(m_{AU})-\gamma B(m_{AU})}{\gamma-1}e^{\frac{k'}{m_{AU}}}
\Bigg\}^{\frac{-1}{2}}\right]  \right)^{-2} - \frac{3}{2},
\end{equation}
\end{widetext}
where $A(m_{AU})=\frac{M_e m_{AU}}{4\pi
\Phi_0}e^{1-\frac{k'}{m_{AU}}}-1$, $B(m_{AU})=\frac{M_e m_{AU}}{4\pi
\Phi_0}e^{1-\frac{k'}{m_{AU}}}+1$, and $W$ is the Lambert (or
product-log) function. From the above realtion, one can realize
that there is a complex valu for $\omega$, for $\gamma >1 $, which
is not acceptable, so $\gamma > 1$ is illegal for this model. Now,
we turn our attention to the values of $\gamma$ that are smaller
than one. In above figure, the $\omega$ parameter and effective gravitational constant have been plotted versus scalar field $\Phi_0$ and mass scale $m_{AU}$, which corresponds to the $r=1 AU=10^8 km$. In Fig.1, $\gamma$-post Newtonian parameter has been taken as $\gamma-1=0.2 \times 10^{-5}$, and $\gamma-1=2.5 \times 10^{-5}$ for Fig.2. In first case, the $\omega$ parameter has been obtained as order of $10^5$, and in the second case, it is in order of $10^4$. As we see, the $\omega$ parameter almost has a constant value for selected range of mass scale and scalar field. These value of $\omega$ satisfy the condition $\omega > 10^4$ {\cite{17-a}} for solar system limit. Also, the effective gravitational constant has been plotted versus the mass scale and scalar field for mentioned values of $\gamma$. For both case, $G_{eff}$ tends to a specific value, which is in order of $10^{-38}$. There is a difference between gravitational constant and effective gravitational constant.

\end{itemize}


\subsection{Second Example}

In this case we set $f(\Phi)=1$, so the model should be back to
the Brans-Dicke model {\cite{17-a}}. From the definition of constant
we realize that $C_1=1$ and $C_2=0$. Like the previous subsection
we investigate this case for $M=0$ and $M\neq 0$.

\begin{itemize}

\item{\bf If $M=0$}\\

In this state, from the definition of constants we have
$$C_2=C_3=k=k'=0.$$ The scalar field get the simple form

\begin{equation}\label {33}
\phi(r)=\frac{M_e}{4\pi(2\omega+3)r},
\end{equation}
and for $h_{\mu\nu}$ we have

\begin{equation}\label {34}
h_{00}=\frac{2(\omega+2)}{(2\omega+3)} \frac{M_e}{4\pi\Phi_0 r},
\end{equation}

\begin{equation}\label {35}
h_{ij}=\delta_{ij} \frac{2(\omega+1)}{(2\omega+3)} \frac{M_e}{4\pi\Phi_0 r}.
\end{equation}
In this case as we expected, $\gamma$-post Newtonian parameter
takes its familiar form {\cite{21}}, namely

\begin{equation}\label {36}
\gamma=\frac{h_{ij}|_{i=j}}{h_{00}}= \frac{\omega+1}{\omega+2},
\end{equation}
and we have effective gravitational constant as

\begin{equation}\label {37}
G_{eff}=\frac{1}{16\pi\Phi_0}\left(
1-\frac{M_e}{4\pi\Phi_0(2\omega+3)r} \right).
\end{equation}

\item{\bf If $M\neq 0$}\\
In this situation, the constant $C_3$ is not vanished, and if we
attend to the action we realize that we have come back to the BD
action, so we should be back to the results of {\cite{17-a}};
however note that we have pick out a different potential
function. For this case, we have
\begin{equation}\label {38}
\phi(r)= \frac{M_e}{4\pi(2\omega+3)r} e^{-kr},
\end{equation}
and also for the component of $h_{\mu\nu}$, we have

\begin{equation}\label {39}
h_{00}=\frac{M_e e^{-k'r}}{4\pi\Phi_0 r} \left( 1+
\frac{e^{-(k-k')r}}{(2\omega+3)} \right),
\end{equation}

\begin{equation}\label {40}
h_{ij}=\delta_{ij} \frac{M_e e^{-k'r}}{4\pi\Phi_0 r} \left( 1-
\frac{e^{-(k-k')r}}{(2\omega+3)} \right).
\end{equation}
From the above relation we can express $\gamma$-post Newtonian
parameter as

\begin{equation}\label {41}
\gamma=\frac{1- \frac{e^{-(k-k')r}}{2\omega+3}}{1+
\frac{e^{-(k-k')r}}{2\omega+3}}.
\end{equation}
And also from (\ref{21}) the effective gravitational constant can
be obtained as

\begin{equation}\label {42}
G_{eff}=\frac{1}{16\pi\Phi_0}\left( 1-\frac{M_e
e^{-kr}}{4\pi\Phi_0(2\omega+3)r} \right),
\end{equation}
when we compare these result with the result of {\cite{17-a}}, we
see that they are in a perfect match, although there is a
difference which appears in $k'$. This difference is resulted
from the sort of potential function we have selected.

\end{itemize}

From above discussion we see that our results is compatible with
the results of previous work {\cite{17-a,21}} and this match can be
a conformation for the validity of Chameleon-Brans-Dicke theory
model.

\section{Conclusion}
In this work, we tried to obtain $\gamma$-post Newtonian parameter in the Chameleon-Brans-Dicke medel. The gravitational constant is promoted to a dynamical scalar field in this model, therefore we could predict the effective gravitational constant as well. In section two, we found out the general form of $\gamma$-post Newtonian parameter, and as we saw it is a
complicated function. In the first part of third section, an exponential function for $f(\Phi)$ has been suggested, and $\gamma$-post Newtonian parameter and effective gravitational constant has been acquired in the presence of potential and without potential either.When we plotted the $\omega$ parameter for two observational values of $\gamma$, we realized that $\omega$ took the values of order $10^4$ which satisfy the solar system constraint. This is an achievement of Chameleon-Brans -Dicke model in solar system limit. Also, we could plot the effective gravitational constant as well. We found out that there is a small difference between gravitational constant and effective gravitational constant. In the last part of this work, we ignored the coupling between matter and scalar field, by setting $f(\Phi)=1$, to consider the compatibility of our resulte with the previous works; after some algebric analysis, we arrived at this result that they are match.\\
The achievements of Chameleon-Brans-Dicke model in larg scale, in description of agegraphic, new egegraphic, and holographic models, and also satisfying solar system constraint indicate advantage of this model.


\end{document}